\documentclass[journal]{IEEEtran}
\newtheorem{theorem}{Theorem}
\usepackage{amsmath,amssymb}
\usepackage{color}

\usepackage{cite}

\ifCLASSINFOpdf
   \usepackage[pdftex]{graphicx}
  \graphicspath{{../pdf/}{../jpeg/}}
   \DeclareGraphicsExtensions{.pdf,.jpeg,.png}
\else
   \usepackage[dvips]{graphicx}
   \graphicspath{{../eps/}}
   \DeclareGraphicsExtensions{.eps}
\fi
\usepackage{amsmath}
\usepackage{algorithmic}

\usepackage{array}
\usepackage{fixltx2e}

\usepackage{stfloats}
\ifCLASSOPTIONcaptionsoff
  \usepackage[nomarkers]{endfloat}
 \let\MYoriglatexcaption\caption
 \renewcommand{\caption}[2][\relax]{\MYoriglatexcaption[#2]{#2}}
\fi
\usepackage{url}

\begin{document}
%
\title{Structure detection of Wiener-Hammerstein systems with process noise}
%
%
%

\author{Erliang~Zhang,
        Maarten~Schoukens,
        and~Johan~Schoukens,~\IEEEmembership{Fellow,~IEEE}
\thanks{This work was supported in part by the Methusalem grant of the Flemish Government (METH-1), by the Belgian Federal Government (IAP DYSCO VII/19), by the ERC advanced grant SNLSID 320378, and by the NSFC Grant U1504618. }
\thanks{E. Zhang is with the Department
of Mechanical Engineering, Zhengzhou University, 450001 Zhengzhou, China (e-mail: erliang.zhang@zzu.edu.cn).}
\thanks{M. Schoukens and J. Schoukens are with the Department of Fundamental Electricity and Instrumentation (ELEC), Vrije Universiteit Brussel, B1050 Brussels, Belgium.}
\thanks{Reference to the final published version: E. Zhang, M. Schoukens and J. Schoukens, Structure detection of {Wiener-Hammerstein} systems with process noise, IEEE Transactions on Instrumentation and Measurement, vol. 66, no. 3, pp. 569-576, March 2017. DOI: 10.1109/TIM.2016.2647418. \textcopyright 
2017 IEEE. Personal use of this material is permitted. Permission from IEEE must be obtained for all other uses, in any current or future media, including reprinting/republishing this material for advertising or promotional purposes, creating new collective works, for resale or redistribution to servers or lists, or reuse of any copyrighted component of this work in other works
}
}
\maketitle

\begin{abstract}

Identification of nonlinear block-oriented models has been extensively studied. The presence of the process noise, more precisely its location in the block-oriented model influences essentially the development of a consistent identification algorithm. The present work is proposed with the aim to localize the process noise in the block-oriented model for accurate nonlinear modeling. To this end, the response of a Wiener-Hammerstein system is theoretically analyzed, the disturbance component in the output, caused by the process noise preceding the static nonlinearity, is shown to be dependent on the input signal. Inspired by such theoretical observation, a simple and new protocol is developed to determine the location of the process noise with respect to the static nonlinearity by using an input signal that is periodic, but nonstationary within one period. In addition, the proposed technique is promising to detect the type of certain static nonlinearity (e.g., dead-zone, saturation). Finally, it is validated on a simulated example and a real-life benchmark. 
\end{abstract}

\begin{IEEEkeywords}
	Structure detection, block-oriented model, process noise, nonstationary input, system identification.
\end{IEEEkeywords}

%
\IEEEpeerreviewmaketitle

\section{Introduction}
%
%
%
%
Block-oriented models are popular in nonlinear system modeling as they are quite simple to understand and easy to use due to the separation of the nonlinear dynamic behavior into linear time invariant (LTI) dynamics and the static nonlinearities \cite{Bai2010}. One most known member of this family is the Wiener-Hammerstein (WH) model, which sandwiches a static nonlinearity between two LTI subsystems, and is popularly used for nonlinear system modeling in biological, electronic and mechanical applications.

Identification methods have been extensively developed for WH systems without the presence of process noise. See, for example, the Special Section in Control Engineering Practice \cite{Hjalmarsson20121095} and the algorithms reported therein, initial estimation \cite{Schoukens2005},  recursive identification \cite{Mu201414}, and {LIFRED \cite{Tan2002}}. Identification of other block-oriented models is also reported, such as cascade\cite{Weiss1998}, parallel Wiener \cite{Schoukens_M2012}, parallel Hammerstein \cite{Schoukens_M2011}, parallel WH system \cite{Schoukens2015111}, and block-structured nonlinear feedback model \cite{Schoukens2008857,Vanbeylen2013}. 

The block-oriented model identification has also been studied with the process noise passing through the static nonlinearity. The inclusion of such disturbing noise significantly complicates the estimation problem, hence an identification framework, which is different from the one applied to the case where the process noise is absent or after the static nonlinearity, should be proposed to obtain the consistent identification of a block-oriented model, such as maximum likelihood identification with the concept of marginalization \cite{Haryanto201354, Hagenblad2008}, expectation-maximization algorithm \cite{Wills2013}, and errors-in-variable formulation \cite{Wahlberg2014}.

To sum up, for the measurement engineer it is extremely important to know if the disturbing process noise is passing through the nonlinearity. If this is the case, the observed noise disturbances at the output of the system are no longer independent of the input. In that case all the classical methods to generate uncertainty bounds fail because these assume explicitly that the disturbing noise is independent of the input. Measurement engineers should be aware of this problem, because in that case alternatives need to be developed to provide reliable uncertainty bounds.

The block-oriented model with the process noise at a wrong location will not be incorrectly fitted to the data. The location of the process noise in the block-oriented model can be viewed as a part of model structure in the sense that it should be assigned prior to implementing any parameter estimation algorithm. Model structure detection is a crucial step in the identification of (block-oriented) nonlinear systems \cite{Haber1990,Beligiannis2005,Lauwers2008,Schoukens2015225}. It is most often performed as part of the measurement process using specific input excitation signals, or frequency response function measurements over a range of setpoints. However, state-of-the-art block-oriented structure detection methods do not guide the user towards the selection of the correct noise framework.

Due to the fact that the disturbance of the process noise is hidden in the measurement noise and nonlinear distortions under general excitations, it is non-trivial to locate the process noise with respect to the static nonlinearity in the block-oriented model based on output observations. The objective of this paper is to tackle this issue by developing a novel measurement protocol, and eventually provide a correct model structure for accurate modeling of nonlinear systems within a complex noise framework.

Besides, the dead-zone and saturation nonlinearities can be present in many nonlinear systems, and often need to be parameterized with high model complexity, the detection of their presence is valuable for model order selection, however it is barely explored without performing system identification.

Taking the WH model as an example, the contributions of the present work are as follows. 
\begin{itemize}
	\item [1)] The output disturbance, contributed by the process noise passing through the static nonlinearity, is explicitly derived as a function of the excitation signal. 
	\item [2)] A structure detection framework for block-oriented models with process noise is established through the use of a specially designed nonstationary input signal.
	\item [3)] {The presence of static nonlinearities of specific types (e.g., dead-zone, saturation) can be detected based on the proposed indicator.}
\end{itemize}

The remainder of the paper is organized as follows. {Section \ref{sec:pf} is dedicated to problem formulation. Section \ref{sec:rwhs} analyzes the response of the WH system. Section \ref{sec:sd} proposes a protocol for structure detection of the WH model in the presence of the process noise, and it is validated numerically and experimentally in Section \ref{sec:exa}. Conclusions are given in Section \ref{sec:con}.}

\section{Problem formulation}
\label{sec:pf}
The following mild assumptions are made on input signal, static nonlinearity, measurement noise and process noise.

\begin{itemize}[\IEEEsetlabelwidth{Z}]
	\item[(A1)] The output measurement noise $e_y(t)$ and the process noise $e_x(t)$ are assumed to be zero-mean stationary noises, mutually uncorrelated, and independent of the input excitation ${u_0}(t)$. {Both disturbing noises can be colored.}
	\item[(A2)] The input signal ${u_0}(t)$ is assumed to be persistent, whose value is bounded.
	\item[(A3)] The static nonlinearity $f(\cdot)$ belongs to the set of generalized nonlinearities described in \cite{Schoukens20081654}. The static nonlinearity function can be approximated arbitrary well by polynomials in the sense that the mean square error tends to zero as the polynomial degree tends to $\infty$. Namely, 
	\begin{equation}
	f(x) {=} \sum_{i=0}^{n_f}a_ix^i.
	\label{eq:static_nl}
	\end{equation}
	where $n_f$ is a positive integer and {tends to $\infty$}.
\end{itemize}

{Towards the real-life measurement and modeling, the present work aims to handle the structure identification of block-oriented models in a complex noise framework where the process noise and measurement noise are both present, the former can simulate any stationary exogenous internal noise. The structure of the block-oriented model is given, while the location of the process noise is unknown. As the process noise is allowed to be colored and $R(q)$ is a linear system, the process noise can equally well be applied anywhere before the nonlinearity with an additional filter. Similarly, it can be equally present anywhere after the static nonlinearity.} Note that the absence of the process noise is a particular example of the case where the process noise succeeds the static nonlinearity, as the identification of the WH model for them is formulated in the same framework.
	
\begin{figure}[bhtp]
	\centering
	\includegraphics[width=0.45\textwidth]{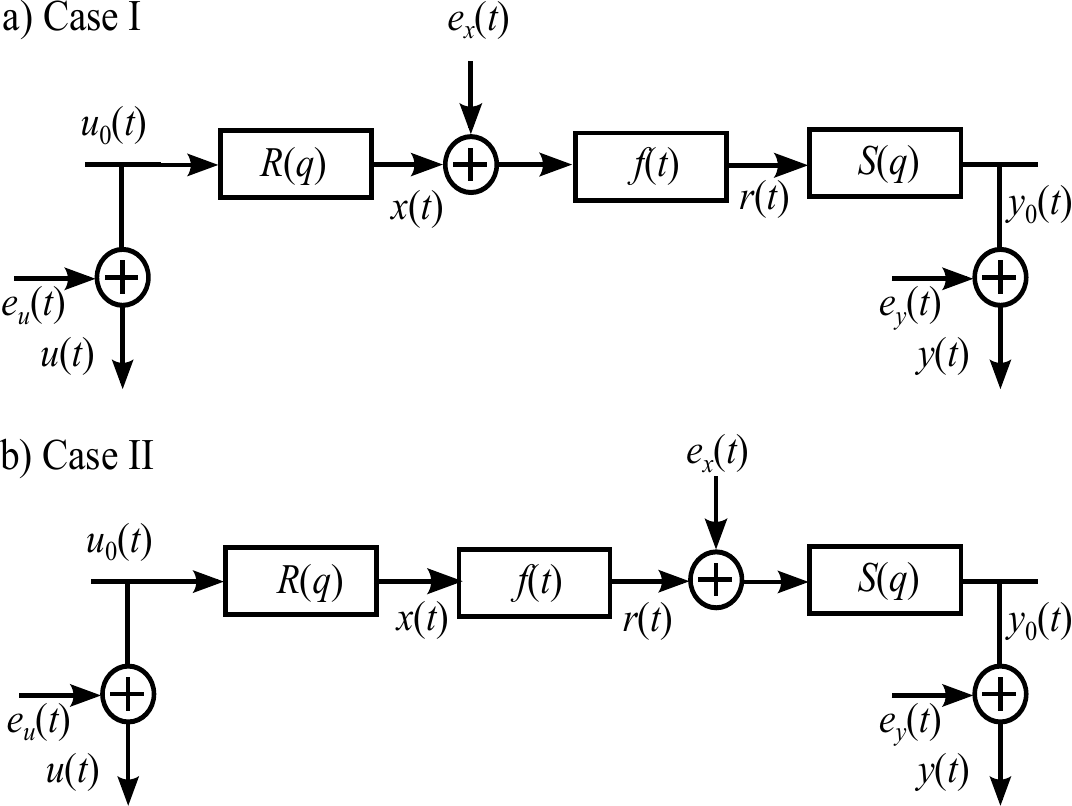}
	\caption{Wiener-Hammerstein models with the process noise. $R(q)$ and $S(q)$ are linear subsystems, $f(t)$ is the static nonlinearity, $u_0(t)$ is the noise-free input, $e_x(t)$ is the process noise, $e_u(t)$ and $e_y(t)$ are the input-output measurement noises, $u(t)$ and $y(t)$ are input-output observations.}
	\DeclareGraphicsExtensions.
	\label{fig:WH}
\end{figure}

Hence, the location of the process noise is considered for the case I and case II, respectively, as shown in Fig. \ref{fig:WH}. {Both of them constitute different patterns of the nonlinear model structure with the process noise. Therefore, the main problem to be solved in this paper is to differentiate these structures of the WH model with the inclusion of the process noise based on output observations only without starting a more demanding system identification process.}

\section{Response of Wiener-Hammerstein system}
\label{sec:rwhs}

\subsection{Process noise passing through the static nonlinearity}
Considering the WH system with the process noise passing through the static nonlinearity, as seen in Fig. \ref{fig:WH} a), it holds that
\begin{align}
\label{eq:x}
x(t) &= R(q)u(t) \\ 
\label{eq:r}
r(t) & = f\left(\tilde{x}(t)\right) \\
\label{eq:y} 
{y_0}(t) & = S(q)r(t) 
\end{align}
where $\tilde{x}(t)=x(t)+e_x(t)$.

Inserting \eqref{eq:static_nl} into \eqref{eq:r}, and then combing \eqref{eq:x} and \eqref{eq:y}, the system response reads
\begin{equation}
y(t) = \sum_{i=0}^{n_f}a_iS(q)x^i(t) + e_p(t) + e_y(t)
\label{eq:y1}
\end{equation}
where the output disturbance caused by the process noise
\begin{align}
e_p(t)  & = \sum_{i=1}^{n_f}\sum_{j=0}^{i-1}a_ic_{ij}S(q)\left\{x^j(t)e^{i-j}_x(t)\right\}
\label{eq:ep1}
\end{align}
with the binomial coefficient
\begin{equation}
c_{ij} = \frac{i!}{(i-j)!j!}
\end{equation}
A toy example is given in what follows to derive explicitly $e_p(t)$ for an {intuitive} understanding. Let $R(q)=S(q)=1$, $f(t)=x^3(t)$, {the input }$u_0(t)$ and {the process noise} $e_x(t)$ are i.i.d. noises with variances $\sigma_u^2$ and $\sigma_e^2$ respectively, 
\begin{align}
y(t) & =(u_0(t)+e_x(t))^3 + e_y(t)  \nonumber \\
& = y_{\text{BLA}}(t) + y_S(t)+e_p(t) + e_y(t)
\end{align}
where the response $y_{\text{BLA}}(t)$ of the underlying linear model, the nonlinear distortion $y_S(t)$, and the output error $e_p(t)$ due to the process noise are written as follows, 
\begin{align}
y_{\text{BLA}}(t) & = 3(\sigma_u^2+\sigma_e^2)u_0(t) \nonumber \\
y_S(t) & =  u_0^3(t)-3(\sigma_u^2+\sigma_e^2)u_0(t) \nonumber \\
e_p(t) & = 3u_0^2(t)e_x(t) + 3u_0(t)e_x^2(t) + e_x^3(t) \nonumber
\end{align}
where {$e_p(t)$ is explicitly written as a polynomial function of $e_x(t)$ with coefficients scaled by $u_0(t)$ and $u_0^2(t)$, and }$y_S(t)$ is found to be free of the process noise $e_x(t)$.

\subsection{Process noise succeeding the static nonlinearity}
The process noise enters into the WH system after the static nonlinearity (see Fig. \ref{fig:WH} b)), it holds in this case that
\begin{align}
\label{eq:r2}
r(t) & = f\left(x(t)\right)\\
\label{eq:yy2} 
y_0(t) & = S(q)\tilde{r}(t) 
\end{align}
where $\tilde{r}(t)=r(t)+e_x(t)$. Combining \eqref{eq:static_nl}, \eqref{eq:x}, \eqref{eq:r2} and \eqref{eq:yy2}, it is easily concluded that 
\begin{equation}
y(t) = \sum_{i=0}^{n_f}a_iS(q)x^i(t) + e_p(t) + e_y(t)
\label{eq:y2}
\end{equation}
where the output disturbance contributed by the process noise
\begin{equation}
e_p(t) =  S(q)e_x(t).
\label{eq:ep2s}
\end{equation}

\subsection{Comments}
\label{sec:comments}
{By analyzing the response of the WH model, the output disturbance caused by the process noise is found to be of different form for the different structures of the WH model shown in Fig. \ref{fig:WH}. The following remarks can be made.}

\begin{itemize}[\IEEEsetlabelwidth{$\gamma$}]
	\item[1)] When the process noise is before the static nonlinearity, $e_p(t)$ is a linear combination of the response of the system $S(q)$ with the excitation of the $e_x^{i-j}(t)$ tuned by $x^j(t)$ at each {time} instant. $e_p(t)$ is therefore dependent on the input signal $u_0(t)$ through the system $R(q)$.
	\item[2)] When the process noise is after the static nonlinearity, $e_p(t)$ is only the response of the system $S(q)$ excited by the process noise $e_x(t)$, which is independent of the input signal $u_0(t)$.
	\item[3)] Although all the derivations of \eqref{eq:ep1} and \eqref{eq:ep2s} are made for the WH model, one can trivially extend the conclusions to other block-oriented models (e.g., Wiener, Hammerstein, parallel Wiener or Hammerstein, block-structured nonlinear feedback model).
\end{itemize}

\section{Structure detection}
\label{sec:sd}
\subsection{Principle}

{The principle of detecting the WH model structure (described in Fig. \ref{fig:WH}) is to exploit the dependence of the output disturbance $e_p(t)$, which is contributed by the process noise, on the input signal. However,} $e_p(t)$ is mixed with the measurement noise and the nonlinear response, and hardly to be separated from them under general random excitation, as seen in \eqref{eq:y1} and \eqref{eq:y2}. {Therefore, the input experiment design is necessary and considered.}

{The class of considered systems is the so-called Wiener system whose input and output have the same periodicity \cite{Pintelon2012}. }A periodic random signal is used, which will lead the component $\sum_{i=0}^{n_f}a_iS(q)x^i(t)$ to be periodic, and separate it from $e_p(t)$ and $e_y(t)$ as the latter terms are nonperiodic. Further, as suggested by \eqref{eq:ep1} {for the pattern in Fig. \ref{fig:WH} a)}, the nonstationarity of the internal variable $x(t)$ can make $e_p(t)$ {nonstationary and hence} distinguishable from the measurement noise $e_y(t)$ by considering the stationarity of the latter, {while $e_p(t)$ computed by \eqref{eq:ep2s} for Fig. \ref{fig:WH} b) is free of $x(t)$ and hence stationary.}

{Therefore, }a periodic random signal, which is however nonstationary within one period, is ideal to be used as an input excitation to detect the structure of WH model with the process noise.

\subsection{Input excitation design}
\label{sec:inputdesign}
The input signal is designed based on the so-called random phase mutltisine, which is defined as follows.
\begin{equation}
	u_0(t) = 1/\sqrt{N} \sum_{k=-N/2+1}^{N/2-1}A_ke^{\jmath\left(2\pi kt/N+\phi_k\right)}
	\label{eq:rpm}
\end{equation}
for $t=1,\cdots,N$ with $N$ the sample number in one signal period, and $\jmath^2=-1$.The phases $\phi_{k}=-\phi_{-k}$ are i.i.d. such that $\mathbb{E}\left\{e^{\jmath\phi_k}\right\}=0$. 

The RPM is by definition periodic, and behaves as Gaussian noise \cite{Pintelon2012}. The amplitude $A_k$ and phase $\phi_k$ of it can be seen as control variables to tune its form into an expected root-mean-square (RMS) envelope. The phases of the RPM are used as design variables while keeping the amplitudes uniformly constant (${A_0}$) in order to satisfy the persistent excitation condition. 

Given an expected RMS envelope that features the nonstationarity of the signal over one period (e.g., steadily increasing and then decreasing amplitude in time with user-defined slopes), the input signal is generated based on the following iterative protocol: 
\begin{itemize}
	\item [1)] At the $i$-th iteration, the input signal $u_0^{i}(t)$ is divided into a series of segments, the instantaneous envelope $ RMS_i(t)$ is obtained by interpolating the RMS values of these segments. The ratio $ RMS_0(t)/RMS_i(t)$, which denotes the difference between the desired shape and that of the $i$-th iteration signal, is used as a scale factor to produce the signal period of the $(i+1)$-th iteration
	\begin{equation}
	u_0^{i+1}(t) = \frac{RMS_0(t)}{RMS_i(t)}u_0^i(t).
	\end{equation}
	\item [2)] Compute the discrete Fourier transform of $u_0^{i+1}(t)$: $U_0^{i+1}(k) = A_{i+1}(k)e^{\jmath\phi_k^{{i+1}}}$, and impose the constraint of the uniform amplitude,
	\begin{equation}
	\tilde{U}_{i+1}(k) = 
	\begin{cases}
	{A_0}e^{\jmath\phi_k^{{i+1}}}     & k \in \mathcal{I}\\
	0  & k \notin \mathcal{I}\\
	\end{cases}
	\label{eq:A}
	\end{equation}
	where $\mathcal{I}$ denotes the excited frequency grid.
	\item[3)] Update $u_0^{i+1}(t)$ by computing the inverse discrete Fourier transform of $\tilde{U}_{i+1}(k)$.
\end{itemize}
Repeat the steps 1) - 3) until the convergence is attained. See \cite{Zhang2010} for the convergence proof of the iterative algorithm. 

\subsection{Bounded output disturbance}
The output disturbance (sum of the measurement noise and disturbance caused by the process noise) can be averaged out from the output observations by using the periodicity. It will be used to define the indicator for the structure of the WH model including the process noise, whose bounded property will be needed and shown in what follows.

\begin{theorem}
Under Assumptions (A1) and (A2), the output disturbance, which comprises $e_y(t)$ and $e_p(t)$ computed by \eqref{eq:ep1} or \eqref{eq:ep2s}, is bounded at each time instant.
\label{theorem_ep}
\end{theorem}
\begin{IEEEproof}
The process noise $e_x(t)$ and the response $x(t)$ of the system $R(q)$ are bounded under Assumptions (A1) and (A2), $x^{j}(t)e^{i-j}_x(t)$ has a finite value as a result. Hence the term $e_p(t)$ in \eqref{eq:ep1} is bounded under the stability constraint of the system $S(q)$. Likewise, it can be easily shown that the term $e_p(t)$ in \eqref{eq:ep2s} is bounded. {The sum of $e_y(t)$ and $e_p(t)$ is also bounded. }
\end{IEEEproof}

\subsection{Proposed indicator}

The nonstationary property of $e_p(t)$ can be {exploited} in principal to determine {the structure of the WH model with the process noise as an internal noise using one experiment}. However, {the nonstationarity of $e_p(t)$ can be invisible by being buried in the (heavy) measurement noise due to the randomness of the input} which is only one realization of the user-defined expected RMS envelope in an experiment (see \eqref{eq:ep1}). {High-order statistics can be a solution for the nonstationarity detection under strict (process and measurement) noise assumption and at the cost of the high computational complexity.} Instead, a multiple-measurement based strategy is {implemented}. $M$ experiments are considered, an input signal $u_0(t)$ with $P$ periods is generated for each experiment using the iterative protocol defined in Section \ref{sec:inputdesign}. 

Under Theorem \ref{theorem_ep}, the variance of the output disturbance (including $e_p(t)$ and $e_y(t)$) exists and is calculated as
\begin{align}
\hat{\sigma}_{e^{[m]}}^2(t) = \frac{1}{P-1}\sum_{p=1}^{P} \left({y}^{[p,m]}(t)-\hat{y}^{[m]}(t)\right)^2
\end{align}
where $\hat{y}^{[m]}(t) = \frac{1}{P}\sum_{p=1}^{P}y^{[p,m]}$. The estimated variance of the output disturbance is further smoothed by using $M$ experiments,
\begin{equation}
\hat{\sigma}_{e}^2(t) = \frac{1}{M}\sum_{m=1}^{M}\hat{\sigma}_{e^{[m]}}^2(t)
\label{eq:sig_e}
\end{equation}

$\hat{\sigma}_{e}^2(t)$ is proposed as an indicator {to detect the model structure shown in Fig. \ref{fig:WH} based on the reasons as follows. 
\begin{itemize}
	\item[1)] \textit{Intuitiveness.} {As is analyzed,} $\hat{\sigma}_{e}^2(t)$ will vary in magnitude within the period due to the nonstationarity of the input signal when the process noise is before the static nonlinearity, while it will remain constant for the model structure with the process noise succeeding the static nonlinearity by considering the noise stationary property.
	\item[2)] \textit{Robustness and simplicity.} The (user-controlled) varying amplitude of the input helps to reveal the presence of the (weak) process noise from the measurement noise when the model includes the process before its static nonlinearity, which is further enhanced by combining multiple realizations of the random input. Weaker noise assumption and heavy measurement noise can be allowed. The proposed indicator is computationally simple as it only involves the second-order statistics. However, this comes at the cost of an increased measurement time. 
	\item [2)] \textit{Type detection of the static nonlinearity.} As the input has a varying amplitude (e.g, gradually increasing from zero then decreasing), the dynamics of $\hat{\sigma}_{e}^2(t)$ within one period can be used to detect the presence of specific static nonlinearities (dead-zone, saturation) when the process noise precedes the static nonlinearity, which also brings useful structure information for nonlinear system modeling and identification.
\end{itemize}
} In addition, the implemented multiple-measurement strategy allows to quantify the nonlinear system {so that the user gets an idea about the transfer function of the plant, the source of the uncertainties and the level of the nonlinearity} \cite{Pintelon2012}.

\section{Examples}
\label{sec:exa}
\subsection{Nonstationary input signal}
\begin{figure}[htbp]
\centering
\includegraphics[width=0.48\textwidth]{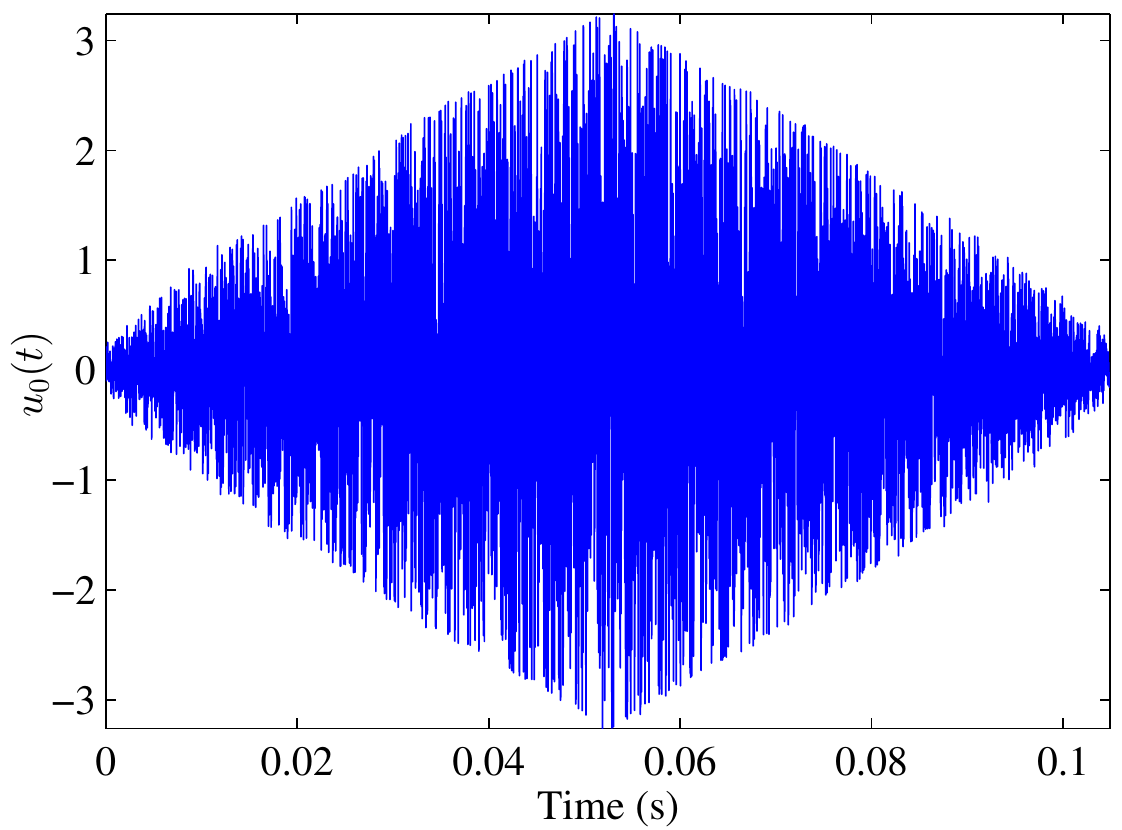}
\caption{Input signal of one period.}
\label{fig:input}
\end{figure}

Setting $A_0=1$ in \eqref{eq:A}, the periodic input signal is generated using the developed iterative protocol from a random realization of the phase of the RPM, as shown in Fig. \ref{fig:input}. It exhibits a steadily increasing then decreasing envelope in time with slopes $\pm 2\sqrt{3}/N$. $N$ is the sample number in one period.

\subsection{Simulated example}

\subsubsection{Plant and noise model}
The LTI subsystems of the WH system under test are described as follows,
\begin{align}
R(z^{-1}) & = \frac{0.1+0.2z^{-1}-0.3z^{-2}}{0.95-1.4z^{-1}+0.9z^{-2}}, \\ 
S(z^{-1}) & = \frac{z^{-1}+0.5z^{-2}}{0.95-0.9z^{-1}+0.9z^{-2}}.
\end{align}
Three typical static nonlinearities are considered, which are polynomial, saturation and dead-zone. The polynomial nonlinearity is described as
\begin{equation}
f\left(x(t)\right)  = \alpha x(t)+\beta x^2(t) + \gamma x^3(t)  
\label{eq:snlpoly}
\end{equation}
with $\alpha = 0.01$, $\beta = 0.02$, $\gamma = -0.008$. 
The saturation nonlinearity is given by 
\begin{equation}
f\left(x(t)\right)	=
\begin{cases}
\alpha_1      & \quad \text{if } x(t)\leq \alpha_1\\
x(t)          & \quad \text{if } \alpha_1<x(t)<\alpha_2\\
\alpha_2      & \quad \text{if } \alpha_2\leq x(t)
\end{cases}
\end{equation}	
with $\alpha_1=-3$ and $\alpha_2=3$. The dead-zone nonlinearity is described by a piecewise-linear function
\begin{equation}
f\left(x(t)\right)	=
\begin{cases}
x(t)-\beta_1      & \quad \text{if } x(t)\leq \beta_1\\
0          & \quad \text{if } \beta_1<x(t)<\beta_2\\
x(t)-\beta_2      & \quad \text{if } \beta_2\leq x(t)
\end{cases}
\end{equation}
with $\beta_1=-1$ and $\beta_2=1$.

{The process noise $e_x(t)$ and $e_y(t)$ are modeled as filtered white noises, which are generated by applying the following filters 
\begin{align}
	H_{e_x}(z^{-1}) & = \frac{1 + 1.8z^{-1}}{1-1.4z^{-1}+0.9z^{-2}} \\
	H_{e_y}(z^{-1}) & = \frac{1 + 2z^{-1} + 5z^{-2}}{1 - 0.94 z^{-1} + 0.88z^{-2}}
\end{align}	
on the i.i.d. white noises with the variances $\lambda_{x}$ and $\lambda_{y}$, respectively.
	}

The sampling frequency is set as 78125 Hz. $N$ = 16384, $M$ = 100, $P$= 100. The structure detection is studied for the two patterns of the model structure illustrated in Fig. \ref{fig:WH}.

\subsubsection{Process noise passing through the static nonlinearity}
\begin{figure}[bhtp]
	\centering
	\includegraphics[width=0.48\textwidth]{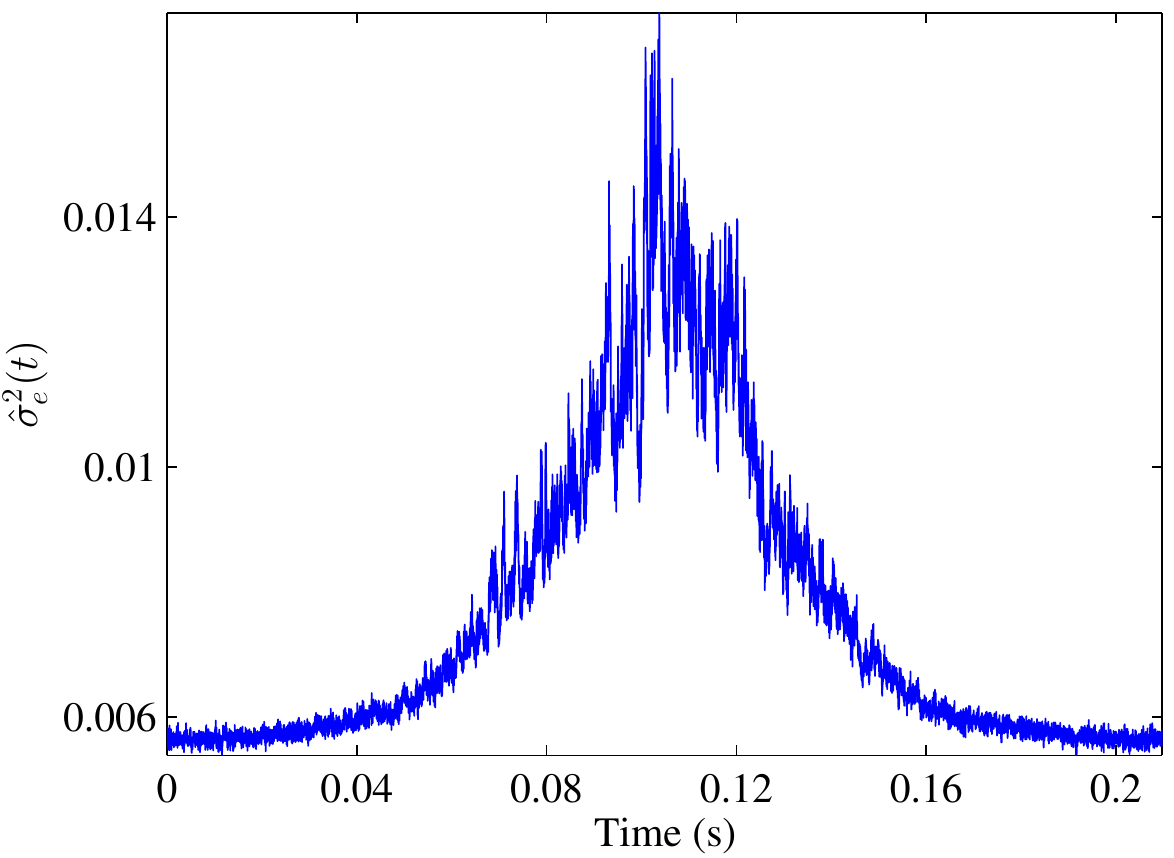}
	\caption{Estimated variance of the output disturbance, $\hat{\sigma}^2_e(t)$, with the 3rd order polynomial nonlinearity which the process noise precedes.}
	\label{fig:poly}
\end{figure}

\begin{figure}[bhtp]
	\centering
	\includegraphics[width=0.48\textwidth]{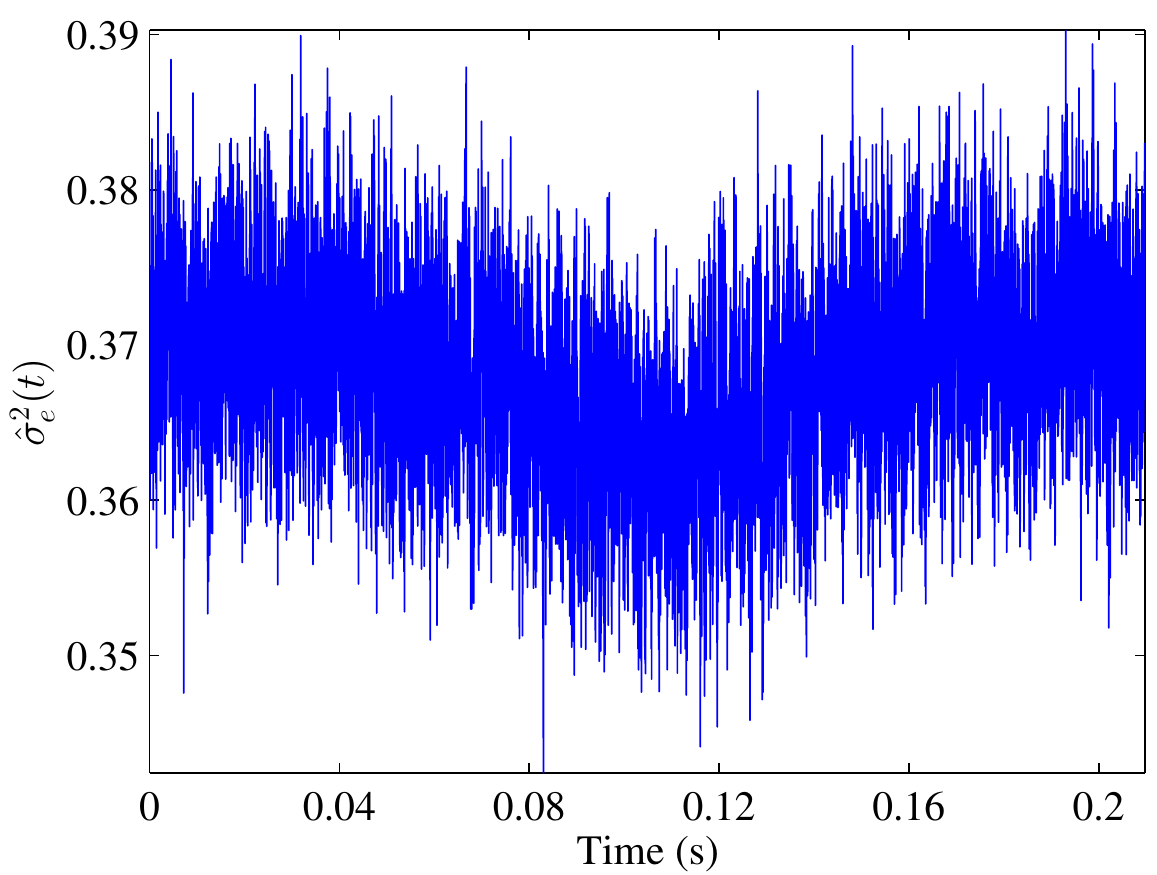}
	\caption{Estimated variance of the output disturbance, $\hat{\sigma}^2_e(t)$, with the saturation nonlinearity which the process noise precedes.}
	\label{fig:sat}
\end{figure}

\begin{figure}[bhtp]
	\centering
	\includegraphics[width=0.48\textwidth]{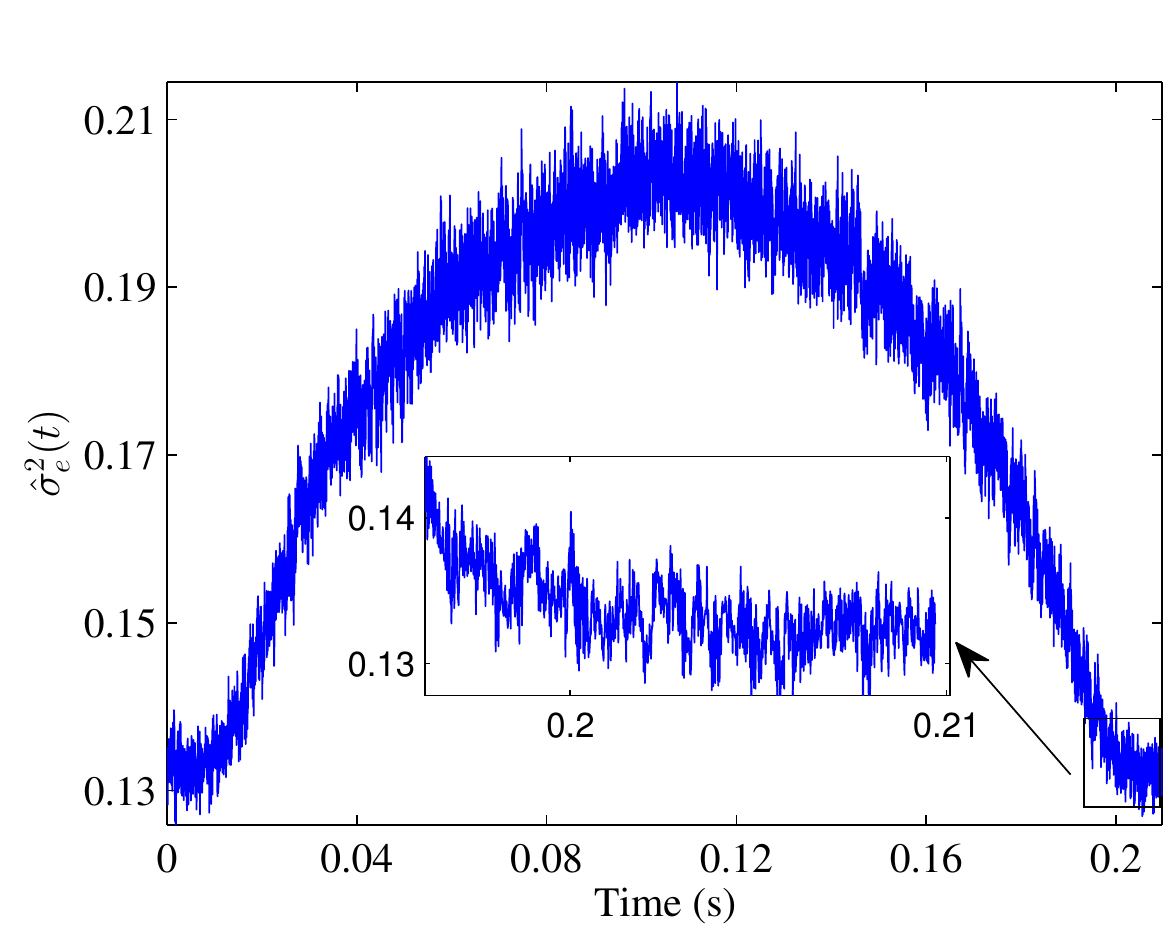}
	\caption{Estimated variance of the output disturbance, $\hat{\sigma}^2_e(t)$, with the dead-zone nonlinearity which the process noise precedes.}
	\label{fig:zone}
\end{figure}

 The process noise level is chosen by setting the SNR ($x(t)$ over $e_x(t)$) as 26 dB, while more measurement output noise is generated to assess the robustness of the proposed technique, the SNR ($y_0(t)$ over $e_y(t)$) equals 20 dB. The variance analysis of the output data, quantified by $\hat{\sigma}^2_e(t)$, is conducted for all three kinds of static nonlinearities, as shown in Figs. \ref{fig:poly} - \ref{fig:zone}. First of all, $\hat{\sigma}^2_e(t)$ is observed to be nonstationary and evolve following the input signal $u_0(t)$ for all these nonlinearities, which is what \eqref{eq:ep1} predicts. Secondly, the form of $\hat{\sigma}^2_e(t)$ can reflect the nature of the static nonlinearity to some extent, as explained in what follows.

{For the used polynomial static nonlinearity, the estimated indicator $\hat{\sigma}^2_e(t)$ evolves gradually, especially varies sightly at the beginning and the end of the period, as illustrated in Fig. \ref{fig:poly}. This is because the system is approximately linear for small amplitude of input signal as the linear term dominates the static nonlinearity (see \eqref{eq:snlpoly}). This property may be used to detect the presence of the linear term in the static nonlinearity.

	}

{For the saturation nonlinearity,} the response of $R(z)$ excited by the input signal $u_0(t)$ attains extreme values around the middle of the period, most of $f(x(t))$ are hence constant and free of the process noise due to the saturation nonlinearity, the response {$y_0(t)$} of $S(z)$ excited by $f(x(t))$ is less contaminated around the middle of the period by the process noise as $S(z)$ is a finite-order system. This explains the behavior of $\hat{\sigma}^2_e(t)$ in Fig. \ref{fig:sat}. 

{For the dead-zone nonlinearity,} the response of $R(z)$ excited by the input signal $u_0(t)$ is weak in magnitude at the beginning and the end of the period, in which $f(x(t))$ are therefore zero in the case of the dead-zone nonlinearity, the response $y_0(t)$ of $S(z)$ is mostly corrupted by the (stationary) measurement noise at both ends of the period. This is in accordance with the observation in Fig. \ref{fig:zone}.

{The following comments are made,
	\begin{itemize}
		\item [1)] The specific dynamics of $\hat{\sigma}^2_e(t)$ over one period can reflect the nature of certain kinds of static nonlinearities (dead-zone, saturation) which the process noise precedes. This is valuable for nonlinear system identification as the absent knowledge on their presence can make the static nonlinearity modeling much involved.
		\item [2)] The choice of $P$, $M$ and the envelope of the input can influence the robustness of the algorithm and the measurement cost. $P$ is preferred to be large enough to average out the output disturbance, multiple experiments ($M>1$) are advised to polish the proposed indicator. $M$ can be reduced by selecting properly the envelope of the input, e.g., increasing its slope for detecting the saturation nonlinearity, while decreasing the slope for the dead-zone nonlinearity.
		\item[3)] The RMS envelope of the input can be further optimized by the user so that more information could be extracted for the static nonlinearity, which will not be elaborated herein.
	\end{itemize}

}
\subsubsection{Process noise succeeding the static nonlinearity}
	
More process noise is introduced, the SNR ($f\left(x(t)\right)$ over $e_x(t)$) equals 20 dB, while less measurement output noise is considered, the SNR ($y_0(t)$ over $e_y(t)$) equals 26 dB. $\hat{\sigma}^2_e(t)$ is found to behave in a stationary manner irrespective of the nonstationary behavior of the input for all three considered static nonlinearities,  as shown in Fig. \ref{fig:SimWH2} a) for the case where the static nonlinearity is the 3rd order polynomial function. This is in agreement with the theoretical result {given by \eqref{eq:ep2s}.  

In addition, the case without the process noise is studied, as shown in Fig. \ref{fig:SimWH2} b). $\hat{\sigma}^2_e(t)$ is only contributed by the measurement output noise, thus smaller than the one estimated with the process noise.

\begin{figure}[!t]
	\centering
	\includegraphics[width=0.48\textwidth]{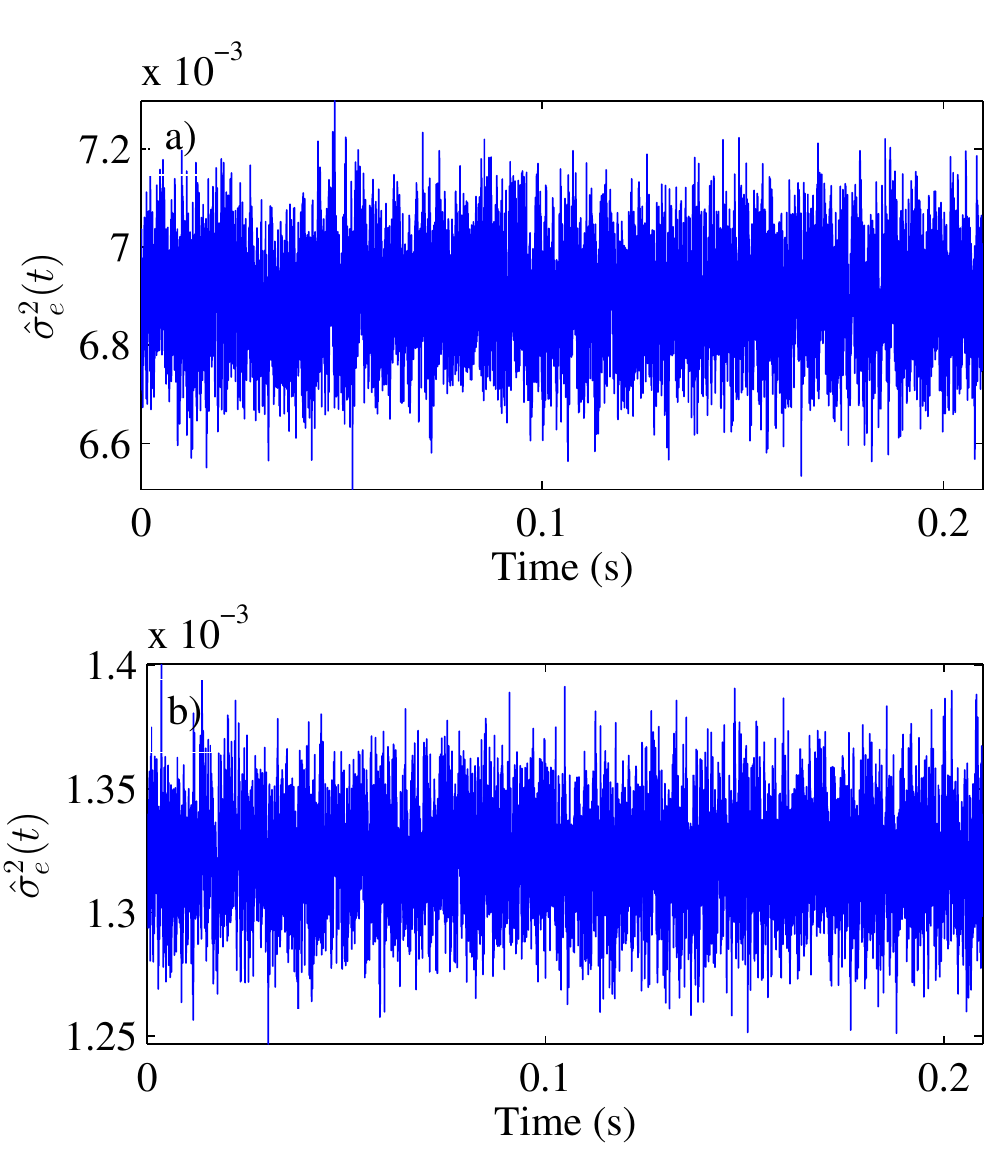}
	\caption{Estimated variance of the output disturbance, $\hat{\sigma}_e^2(t)$. a) with the process noise succeeding the 3rd order polynomial nonlinearity, b) without the process noise.}
	\label{fig:SimWH2}
\end{figure}

\subsection{Experimental example}

\subsubsection{Benchmark setup}
The experimental example is a WH benchmark which contains dominant process noise \cite{Schoukens2016}. The first filter $R(q)$ can be described well with a third order low pass filter. The second LTI subsystem $S(q)$ is designed as an inverse Chebyshev filter with a stop band attenuation of 40 dB and a cutoff frequency of 5 kHz. The second LTI subsystem has a transmission zero within the excited frequency range. The static non-linearity is realized with a diode-resistor network (see Fig. \ref{fig:SNL}), which results in a saturation nonlinearity. 
	
The inputs and the process noise are generated by an arbitrary zero-order hold waveform generator (AWG), the Agilent/HP E1445A, sampling at 78125 Hz. The generated zero-order hold signals are passed through a reconstruction filter (Tektronix Wavetek 432) with a cut-off frequency of 20 kHz. The in- and output signals of the system are measured by the alias protected acquisition channels (Agilent/HP E1430A) sampling at 78125 Hz. The AWG and acquisition cards are synchronized with the AWG clock, and hence the acquisition is phase coherent to the AWG. Leakage errors are hereby easily avoided. Finally, buffers are added between the acquisition cards and the in- and output of the system to avoid that the measurement equipment would distort the measurements. See \cite{Schoukens2016} for  more information.
	
\begin{figure}[bhtp]
	\centering
	\includegraphics[width=0.41\textwidth]{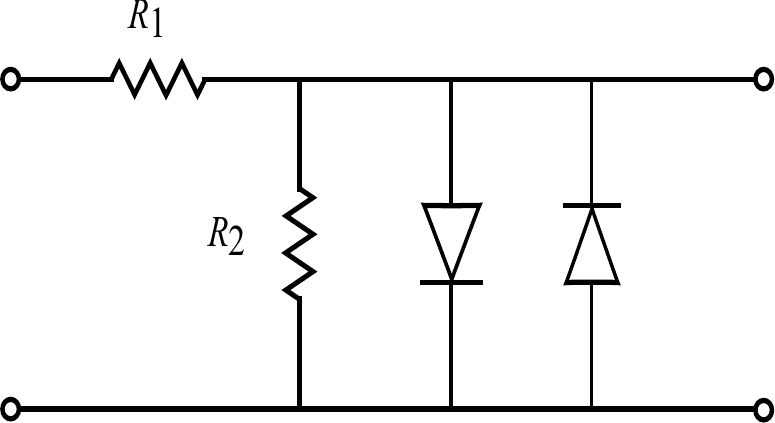}
	\caption{The circuit used to generate the saturation nonlinearity.}
	\label{fig:SNL}
\end{figure}

{In order to validate the proposed approach, the location of the process noise in the benchmark is considered, respectively, for three cases. 1) The process noise enters into the benchmark before the diode-resistor network, as shown by the case I in Fig. \ref{fig:WH}. 2) The process noise, which has the same statistical properties as the one used for 1), disturbs the output of the benchmark system, this noise location is completely equivalent to the case II in Fig. \ref{fig:WH} with an additional linear filter. 3) The process noise is absent, only a bit of noise in the measurement channel is present.} $N=2048$. $M = 32$, $P=32$. The first period of the output data is discarded to reduce the transient effect.

\subsubsection{Process noise passing through the saturation nonlinearity}
The estimated $\hat{\sigma}_{e}^2(t)$ is found to vary within the period, as shown in Fig. \ref{fig:ExpWH}, which suggests that the process noise passes through the static nonlinearity. Moreover, the shape of $\hat{\sigma}_{e}^2(t)$, which is similar to the one in Fig. \ref{fig:sat}, implies the existence of the saturation nonlinearity (generated by the current-voltage characteristic of the diode circuit). This is in compliance with the experimental setup. 

\begin{figure}[bhtp]
	\centering
	\includegraphics[width=0.48\textwidth]{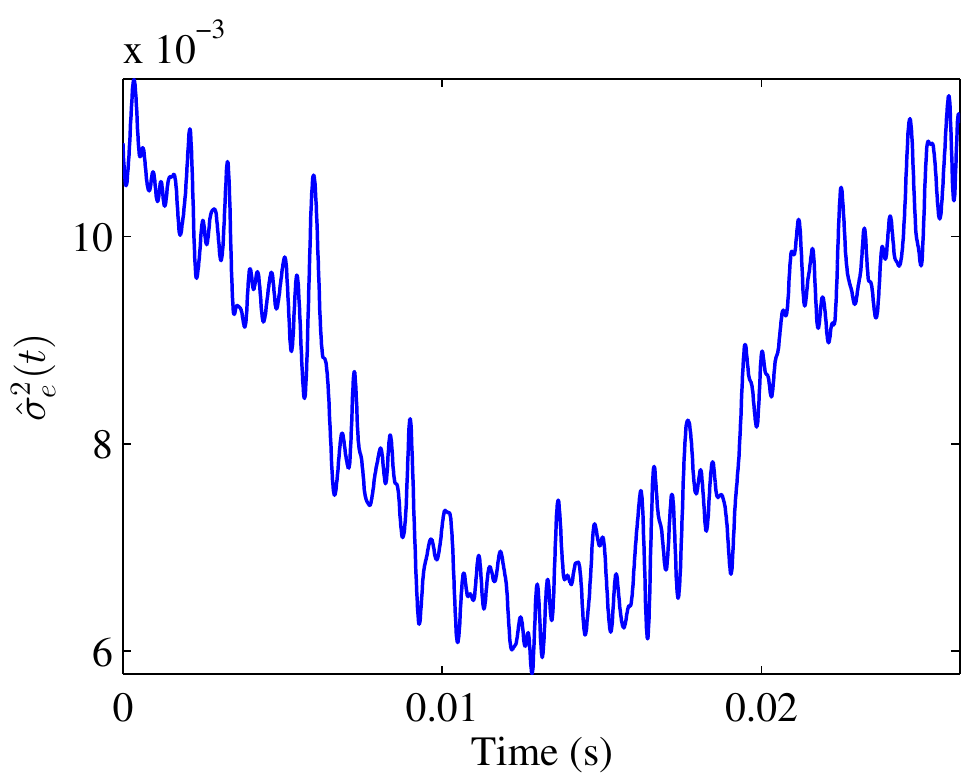}
	\caption{Estimated variance of the output error, $\hat{\sigma}_e^2(t)$, in one period with the process noise passing through the saturation nonlinearity.}
	\label{fig:ExpWH}
\end{figure}

\subsubsection{Process noise succeeding the saturation nonlinearity}

As expected, $\hat{\sigma}^2_e(t)$ is shown to behave in a stationary manner irrespective of the nonstationary behavior of the input, as shown in Fig. \ref{fig:ExpWH_nn} a). This fully conforms with the theoretical and simulated results. Also, the result for the case without the process noise is illustrated in Fig. \ref{fig:ExpWH_nn} b).

\begin{figure}[bhtp]
	\centering
	\includegraphics[width=0.48\textwidth]{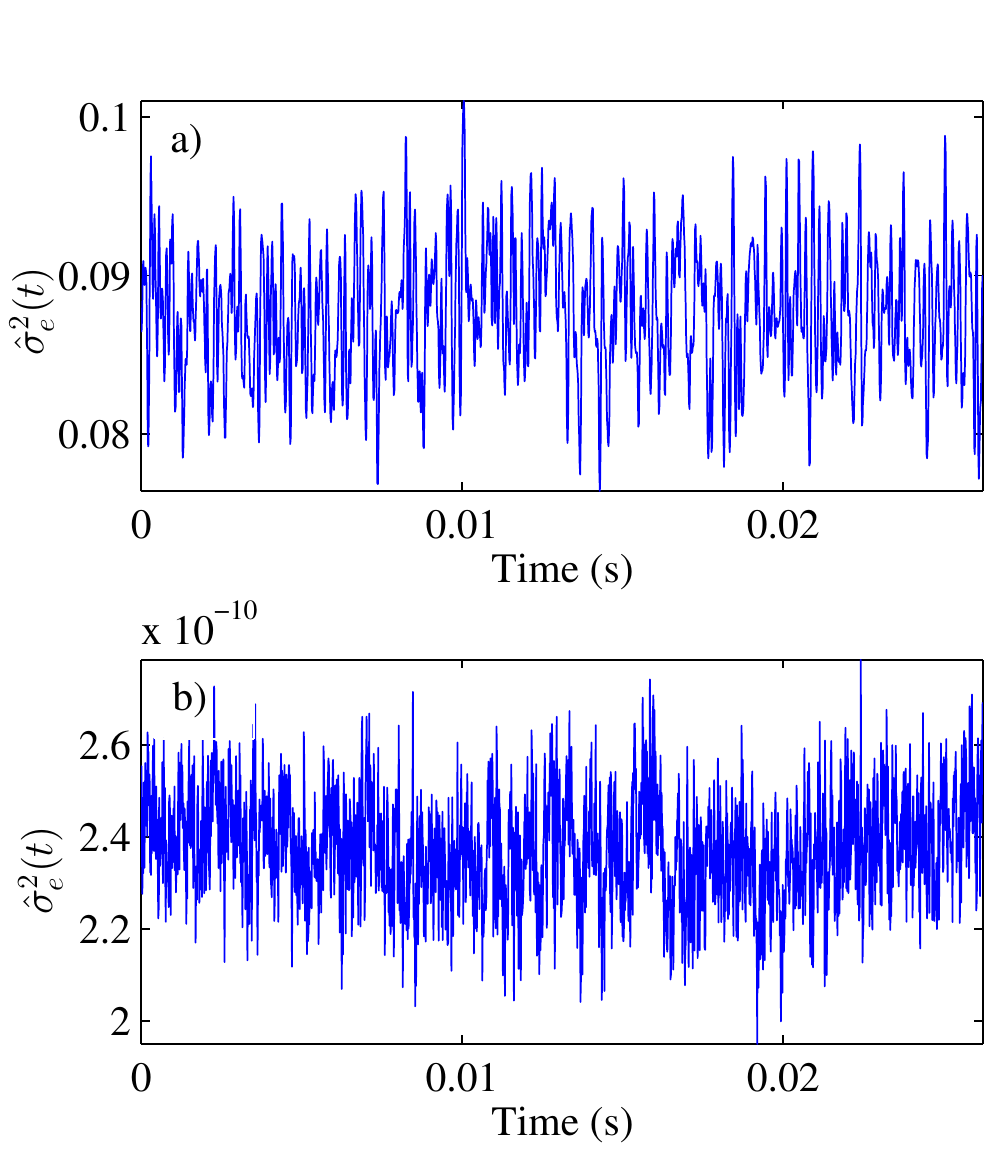}
	\caption{Estimated variance of the output disturbance, $\hat{\sigma}_e^2(t)$. a) with the process noise succeeding the saturation nonlinearity, b) without the process noise.}
	\label{fig:ExpWH_nn}
\end{figure}

\section{Conclusion}
\label{sec:con}
In this paper, a user-friendly technique has been proposed for the structure {detection} of Wiener-Hammerstein systems with the inclusion of the process noise, and {whose effectiveness and robustness are demonstrated} on a simulated example {by considering several kinds of static nonlinearities}, and validated on a real-life Wiener-Hammerstein benchmark. The proposed technique is found to be able to {detect the presence of certain specific static nonlinearities (e.g., dead-zone, saturation)} which the process noise precedes, and it can be {directly} applied to other block-oriented models. {Thus, it can be seen as a useful tool to detect the block-oriented model structure for accurate modeling within a complex noise framework.}

\end{document}